\documentclass[12pt,preprint]{aastex}
\usepackage{amsmath}

\begin{document}
 
\def\etal   {{\it et al. }}                     
\def\msol   {\hbox{M$_\odot$}}                  
\def\ee #1 {\times 10^{#1}}          
\def\ut #1 #2 { \, \textrm{#1}^{#2}} 
\def\un #1 { \, \textrm{#1}}          
\def\u #1 { \, \textrm{#1}}          
\def\kms {\,\textrm{km\,s}^{-1}}
\def\percc {\,\mathrm{cm}^{-3}}

\newcommand\rpc{r_\mathrm{pc}}
\newcommand\msun{\mathrm{M_\odot}}

\shorttitle{Megamaser Disks and Black Holes in AGN}
\shortauthors{Wardle \& Yusef-Zadeh}

\title{A Scaling Relation Between Megamaser Disk Radius and Black Hole Mass in Active Galactic Nuclei}
\author{Mark Wardle$^1$ and Farhad Yusef-Zadeh$^2$}
\affil{$^1$ Astronomy, Astrophysics \& Astrophotonics Research Centre and \\ Department of Physics \& Astronomy, Macquarie University, Sydney, NSW 2109, Australia; mark.wardle@mq.edu.au}
\affil{$^2$Department of Physics \& Astronomy, Northwestern University, Evanston, IL 60208; zadeh@northwestern.edu} 


\begin{abstract} 

Several thin, Keplerian, sub-parsec megamaser disks have been discovered in the nuclei of active galaxies and used to precisely determine the mass of their host black holes.   We show that there is an empirical linear correlation between the disk radius and the black hole mass.    We demonstrate that such disks are naturally formed by the partial capture of molecular clouds passing through the galactic nucleus and temporarily engulfing the central supermassive black hole.  Imperfect cancellation of the angular momenta of the cloud material colliding after passing on opposite sides of the hole leads to the formation of a compact disk.  The radial extent of the disk is determined by the efficiency of this process and the Bondi-Hoyle capture radius of the black hole, and naturally produces the empirical linear correlation of the radial extent of the maser distribution with black hole mass.  The disk has sufficient column density to allow X-ray irradiation from the central source to generate physical and chemical conditions conducive to the formation of 22\,GHz H$_2$O masers.    For initial cloud column densities $\la 10^{23.5}\ut cm -2 $ the disk is non-self gravitating, consistent with the ordered kinematics of the edge-on megamaser disks; for higher cloud columns the disk would fragment and produce a compact stellar disk similar to that observed around Sgr A* at the galactic centre. 
\end{abstract}

\keywords{accretion, accretion disks --- galaxies: Seyfert --- Galaxy: center --- ISM: clouds --- masers --- stars: formation}

\section{Introduction}
\label{introduction} 

The nuclei of some Seyfert 2 galaxies are home to powerful 22\,GHz water masers located in a circumnuclear disk within a parsec of the central massive black hole. In almost edge-on systems, VLBI observations of the maser kinematics enable an accurate determination of the black hole mass.  The archetypal system is NGC 4258, for which the black hole mass has been measured to 1\% accuracy (e.g., Herrnstein et al.~2005; Humphreys et al.~2008).   Recent VLBA observations of Seyfert 2 galaxies have discovered a limited number of additional examples with high inclination angles, allowing precise measurements of the mass of their host black holes as well as the physical size of the maser disks, which range from 0.1--1\,pc (Herrnstein et al.\ 2008; Kuo et al.~2010).

The physical parameters of the disks can be inferred from the conditions necessary to generate the masers.  Collisional inversion of the 22\,GHz H$_2$O transition requires densities of $10^7$--$10^{11}\percc$, temperatures in the range 300--1\,000\,K, and a sufficient column of water to ensure maser amplification (e.g.~Neufeld \& Melnick 1991). These conditions are plausibly produced by irradiation of a molecular disk by the central X-ray source as long as the surface density exceeds $\sim 1\un g \ut cm -2 $ and warping exposes the disk surface to the center (Neufeld et al.~1994; Maloney 2002).   Parsec-scale maser disks must therefore have masses $\ga 10^4$\msol.  

Further insight is provided by examining the center of the Milky Way Galaxy, where there is strong evidence of star formation occurring in a sub-parsec scale disk $\approx 6\times 10^6$ years ago (Paumard et al.~2006).  Approximately 100 massive stars orbit within a few tenths of a parsec of the $\sim 4\times10^6\msol$ black hole Sgr A*  (Bartko et al. 2010; Lu et al. 2010; Do et al. 2009; see also the recent review by Genzel et al. 2010 and references cited therein).  About 2/3 of these are localized in a clockwise rotating stellar disk with a wide range of eccentricities (Levin \& Beloborodov 2003), with the remainder loosely distributed, possibly in a larger counter-rotating disk (Paumard et al. 2006; but see Lu et al.~2009).  Stellar disks could be created by the tidal disruption of an inspiralling stellar cluster (Gerhard 2001; McMillan \& Portegies Zwart 2003; Portegies Zwart et al. 2003; Kim et al. 2004; G\"urkan \& Rasio 2005), but the modelling implies that this mechanism produces a far more disordered stellar system than observed, and a non-existent population of massive stars shed from the cluster extending beyond 0.3 pc from Sgr A*; furthermore the inspiralling time scale is longer than the stellar ages (Paumard et al. 2006; but see Fujii et al. 2008).    A more attractive alternative is that these disks could form ``in-situ'' by gravitational collapse in a disk of gas captured by the black hole (Levin \& Beloborodov 2003; Nayakshin et al.\ 2007), a process previously considered in the context of AGN (Kolykhalov \& Sunyaev 1980; Shlosman \& Begelman 1987; Collin \& Zahn 1999; Goodman 2003). 

The compactness of the stellar disks relative to molecular cloud dimensions implies that the raw material for the  disk is captured from a cloud as it temporarily engulfs Sgr A* while passing through the central parsec of the Galaxy (c.f.\ Sanders 1981; Bottema \& Sanders 1986); the cancellation of angular momentum of the captured cloud material that passes on opposite sides of the black hole naturally produces a compact, gravitationally unstable disk that is consistent with dimensions, mass and kinematic properties of the observed stellar disk (Wardle \& Yusef-Zadeh~2008, hereafter Paper I; Bonnell \& Rice~2008; Mapelli et al.~2008; Alig et al.~2011).  This process may also be responsible for the parsec-scale circumnuclear ring (Sanders~1998; Paper I).

A link between the stellar disk at the Galactic center and AGN megamaser disks was suggested by Milosavljevi\'c \& Loeb (2004) who argued that the masing disks are gravitationally unstable and would eventually create stellar disks analogous to those seen around Sgr A*.    However, the maser distribution and kinematics in the edge-on systems are very thin and close to keplerian,  so that at least in these systems the megamaser disks are gravitationally stable.

In this \emph{Letter}, we show that the cloud capture scenario outlined in Paper I creates gravitationally stable sub-parsec disks with physical conditions conducive to the formation of megamasers.  We begin by showing that systems with well-determined host black hole masses and Keplerian rotation profiles display a strong correlation between disk size and black hole mass (\S2).  In \S 3 we consider a simple model for the partial capture of molecular clouds and demonstrate that this correlation is a natural outcome of the cloud-engulfment scenario, and that, for reasonable parameters, initial cloud column densities $\sim 10^{23}\ut cm -2 $ will create a non-self-gravitating molecular disk consistent with the observed megamaser disks. Our conclusions are summarised in \S4.

\section{A Correlation Between Megamaser Disk Size and Black Hole Mass}

Fourteen resolved parsec-scale megamaser disks in AGN have been reported in the literature.   Of these, three disks (in NGC 1068, NGC 3079, and IC 1481) have maser distributions that are spatially and kinematically disordered, with rotation curves that are flatter than Keplerian (Greenhill et al.~1996; Kondratko et al.~2005; and Mamyoda et al.~2009, respectively).  This is suggestive of possibly clumpy material distributed within a dynamically hot torus with the gravitational potential  dominated by the torus material or a central stellar cluster.   Three others cases (NGC 3393, NGC 4945 and Circinus) have poorly-determined rotation curves so that the black hole mass and dynamical state of the disk are uncertain (Kondratko et al.~2008; Greenhill et al.~1997, 2003 respectively).

As our model for the dynamics of disk formation relies on the black hole mass dominating the gravitational potential on a  scale several times that of the disk (see \S3), we focus our attention on the remaining eight megamaser disks with well-determined, Keplerian profiles and accurate black hole masses (Herrnstein et al.\ 2008; Kuo et al.~2011).   In each case the upper limits to the disk mass estimated from the errors to the fit to the rotation profile are a few percent of the black hole mass. Table 1 lists the the measured mass of the black hole, and the inner and outer radii of the disk as traced by the masers  for each of these eight megamaser disks.  For comparison we have also included the range of radii of the orbits of S-stars forming the compact stellar disk around Sgr A* (Lu et al.~2010).  The remarkable similarity of size scale to the megamaser disks suggest that they may have been formed the same way.
	
\begin{deluxetable}{lrr}
\tablecaption{Physical Characteristics of Megamaser Disks}
\tablewidth{0pt}
\tablehead{
\colhead{Source} &
\colhead{BH Mass 10$^7$ \msol } &
\colhead{Disk Size (pc) }
}
\startdata
NGC 1194$^a$  & 6.5$\pm$0.3 &  0.53-1.33     \\
NGC 2273$^a$  & 0.75$\pm$0.04 & 0.028-0.084  \\
UGC 3789$^a$  & 1.04$\pm$0.05 &  0.084-0.30  \\
NGC 2960$^a$  & 1.16$\pm$0.05 &  0.13-0.37   \\
NGC 4258$^b$  & 3.7$\pm$0.01 &  0.17-0.29    \\
NGC 4388$^a$  & 0.84$\pm$0.02 &  0.24-0.29   \\
NGC 6264$^a$  & 2.91$\pm$0.04 &  0.24-0.80   \\
NGC 6323$^a$  & 0.94$\pm$0.01 &  0.13-0.30   \\
\\
Sgr A*$^c$  & 0.43$\pm$0.02 &  0.038-0.13   \\

\enddata
\tablenotetext{a}{Kuo et al.~2011}
\tablenotetext{b}{Herrnstein et al.~2008}
\tablenotetext{c}{compact stellar disk; Lu et al.~2008}
\end{deluxetable}

For each of these disks we plot the range of disk radii traced by megamasers against black hole mass (see Fig.\ 1).   Remarkably, there is a well-defined upper envelope to the  disk radii that scales linearly with black hole mass, given approximately by 
\begin{equation}
    R_\mathrm{max} \approx 0.3\,\,M_7\;\mathrm{pc} \,,
\end{equation}
where the central black hole mass is $M=M_7\times10^7$\,M$_\odot$.
The outer edge of the maser distribution may simply trace where the surface density of the disk falls below the $\sim 1$\,g\,cm$^{-2}$  needed to shield molecules from the hard X-ray flux from the central source  (Maloney 2002).   However, there is no obvious reason why this radius should follow this trend.  Our preferred explanation, outlined in the following section,  is that the outer radii indicate a genuine physical truncation, and that the linear dependence on black hole mass reflects the kinematics of cloud capture process that forms the disk (see eq.~5). 

\begin{figure}
\center
\includegraphics[scale=1,angle=0]{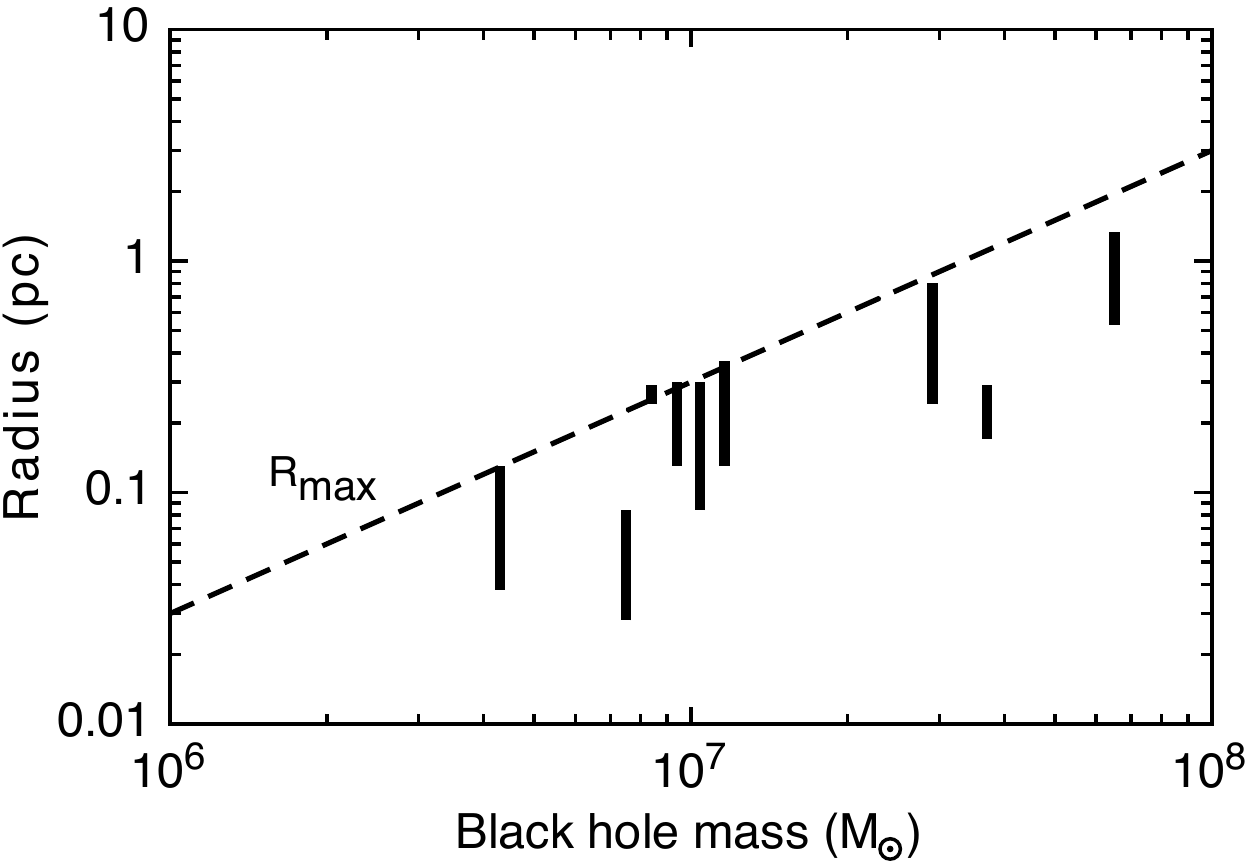}
\caption{
Vertical bars indicate the range of radii of the maser distribution for published H$_2$O megamaser disks with well-determined Keplerian rotation curves versus central black hole mass (Kuo et al.\ 2011; Herrnstein et al.\ 2005) .  The extent of the compact stellar disk at the Galactic center (lowest-mass point; Lu et al.\ 2008) is plotted for comparison. The errors in the black hole masses and radii are a few percent.  The dashed line represents an approximate empirical upper envelope to the outer radii given by 
$R_\mathrm{max} = 0.3 \, (M/10^7\msun)$\,pc (see text). } 
\end{figure}

\section{Disk Formation by Cloud Capture}

In Paper I we showed that partial capture of molecular clouds passing through the central parsec of the galaxy by the $4\times 10^6$\,\msol\ black hole Sgr A* would  naturally create a disk on the 0.1 parsec scale of the stellar disk.  There we focussed on the formation of self-gravitating disks, which because of the short cooling time of the gas, create stars on a timescale comparable to a few orbital times, so that star formation would occur before the disk settles into a thin ordered structure; the resulting stellar orbits have a range of eccentricities and inclinations.

Here we apply this scenario to the very thin and well-ordered megamaser disks, with two key differences from our earlier analysis.  First, we explicitly follow the scaling with black hole mass rather than fixing it at $\approx4\ee 6 \msun $ appropriate for  Sgr A*;  second, we focus on capture events that produce non-self-gravitating disks, which therefore settle into very thin Keplerian disks rather than form stars.

Consider, then, an extended cloud with hydrogen column density $N_\mathrm{H}=N_{23}\ee 23 $\,cm$^{-2}$ that enters the black hole's sphere of influence with speed $v = 225\,v_{225}\kms$. On its passage through the region, the cloud temporarily engulfs the black hole and shocks induced by gravitational focussing of the material passing close to the hole lead to capture of some material, which circularizes.  Rapid cooling of the gas leads to formation of a thin, Keplerian disk.

In this scenario the disk mass is determined by the mass of gas with impact parameter inside the Hoyle--Lyttleton radius (Hoyle \& Lyttelton 1939)
\begin{equation}
    b_0 = \frac{2GM}{v^2} = 1.7\,\frac{M_7}{v_{225}^2}\;\un pc\,,
    \label{eq:b0}
\end{equation}
and its size is determined by the imperfect cancellation of the angular momentum of the captured material that approached on opposing sides of the black hole due to the cloud morphology and structure.  As in Paper I we characterize the uncertain capture dynamics using two parameters.  The first of these, $\kappa$, is the ratio of the captured mass to the Hoyle--Lyttleton estimate, allowing us to write the disk mass as
\begin{equation}
	M_d = \kappa \; \pi  b_0^2 \; (1.4 \, m_\mathrm{H} N_\mathrm{H}) = 1.0\times10^4 \; \frac{\kappa N_{23}M_7^2}{v_{225}^4}\;\;\msun\,.
	\label{eq:Md}
\end{equation}
where $m_\mathrm{H}$ is the mass of a hydrogen atom and we have assumed the standard helium to hydrogen of 0.4 by mass.  We expect that $\kappa \sim 1$ because bulk kinetic energy is efficiently lost through shocking and rapid radiative cooling during the collision (e.g.\, Edgar 2004, and references therein).

The second parameter, $\lambda$, is the average ratio of the specific angular momentum of a fluid element in the resulting Keplerian disk, $(GMr)^{1/2}$, to its initial angular momentum $bv$.  This ignores a number of complications: the cancellation of angular momentum is sensitive to the distribution of inhomogeneities in the cloud and will be highly variable; it will also depend on the orientation of a fluid elements trajectory relative to the final disk plane.  While fluid elements with the same impact parameter $b$ but with different specific angular momentum vectors will end up on different disk radii, the convolution in mapping the initial impact parameter $b$ to eventual location in the disk $r$ also has the effect of averaging out these differences.  This assumption allows us to relate the impact parameter of a fluid element in the approaching cloud, to its typical location in the disk once it has been captured and its motion circularized:
\begin{equation}
	r = \frac{\lambda^2 v^2 b^2}{GM} = \frac{2\lambda^2 b^2}{b_0}\,.
	\label{eq:r-vs-b}
\end{equation}
The value of $\lambda$ is uncertain, but recent simulations by Alig et al.\ (2011) suggest that $\lambda\sim 0.3$ -- $0.4$.  

The disk radius is determined by the initial angular momentum of fluid elements with impact parameter $b_0$, for which eq (\ref{eq:r-vs-b}) yields 
\begin{equation}
	R_d = 2\lambda^2 b_0 \approx 0.31\,\frac{\lambda_\mathrm{0.3}^2M_7}{v_{225}^2}\;\;\mathrm{pc} \,,
	\label{eq:Rd}
\end{equation}
and so the mean surface density of the disk is
\begin{equation}
    \Sigma_d \equiv \frac{M_d}{\pi R_d^2} = \frac{\kappa}{4\lambda^4}\,(1.4\, m_\mathrm{H} N_\mathrm{H}) \approx 7.2 \; \frac{\kappa N_{23}}{\lambda_{0.3}^4}\;\u g \ut cm -2  \,.
    \label{eqn:Sigmad}
\end{equation}
Strikingly, the disk size corresponds to the upper envelope of the maser disks with the canonical model parameters $v \approx 225 \kms$ and $\lambda\approx 0.3$.  In other words, the kinematics of the model naturally produces the observed extent of the maser disks, \emph{and} the empirical scaling with black hole mass.  

To estimate the surface density profile of the disk we note that eq (\ref{eq:r-vs-b}) implies that a thin cylindrical ring of cloud material with impact parameters in the range $[b,b+db]$ ends up in the annulus $[r,r+dr]$ in the disk.  Then mass conservation relates the cloud and disk surface densities via $2\pi b\,\kappa(1.4 \, m_\mathrm{H} N_\mathrm{H})\,db\, = 2\pi r\,\Sigma(r)\,dr$ implying that the disk has a $1/r$ surface density profile:
\begin{equation}
	\Sigma(r) =   \lambda^2\Sigma_d \, \frac{b_0}{r}
   \approx 1.1\; \frac{\kappa N_{23}}{\lambda_{0.3}^2\, v_{225}^2}\,\frac{M_7}{\rpc}\;\un g \ut cm -2 \,.
	\label{eq:Sigma-profile}
\end{equation}
where in the final expression $\rpc$ is the cylindrical radius in parsecs. For this profile the disk mass enclosed within radius $r$ of the BH is linearly proportional to $r$, and the density at the disk edge is $ 0.5\,\Sigma_d$.

This simple model disk must meet three requirements in addition to its size to guarantee that masers may be present over the radial extent of the disk.  First, the mass given by eq (\ref{eq:Rd}) must not exceed the mass of the incoming cloud, otherwise the disk will be truncated at a smaller radius where the enclosed disk mass equals the cloud mass $M_\mathrm{cl}$.  

Second, collisional inversion of the 22\,GHz H$_2$O transition requires densities of $10^7$--$10^{11}\percc$, temperatures in the range 300--1\,000\,K, and a sufficient column of water to ensure significant maser amplification (e.g.~Neufeld \& Melnick 1991). These conditions are produced by X-ray irradiation of the disk as long as warping exposes the disk surface to the central X-ray source and the disk column density exceeds $\sim 1\un g \ut cm -2 $  (Neufeld, Maloney \& Conger 1994; Maloney 2002).  For lower column densities the X-rays dissociate any molecules in the disk, for higher column densities there is a transition from atomic gas to a water-bearing molecular layer layer with temperatures of $\sim 400\,$K.   Because inversion of the 22\,GHz line can occur in this layer for a broad range of densities, the maser action is insensitive to the X-ray flux and disk column.  Therefore we need only adopt as a requirement that the column density of our model disks exceed $1\u g \ut cm -2 $.   Third, the disk should be gravitationally stable for $r\la R$, i.e. $\Sigma(r)\la c_s\Omega/\pi G$, where $c_s$ is the isothermal sound speed at $T\approx 400$\,K and $\Omega$ is the Keplerian orbital frequency. 
These constraints are most stringent at $R_d$, and for our $1/r$ surface density profile the density at the disk edge is $ 0.5\,\Sigma_d$,  yielding the limits $\Sigma_d \la M_\mathrm{cl}/(\pi R_d^2)$ and 1\,g\,cm$^{-2} \la \frac{1}{2} \Sigma_d \la c_s \Omega(R_d) /\pi G$.  

The constraints on $\Sigma_d$ are mapped to the cloud column density using eq (\ref{eqn:Sigmad}) with $v=225\kms$, $\kappa=1$ and $\lambda=0.3$ and are plotted in Figure~2 as a function of black hole mass.  To recap, the column of the incoming cloud must lie within the unshaded region to guarantee that the disk formed by the captured material is thick enough to allow the X-rays to provide the physical conditions and water column needed to produce the megamaser emission, but not so thick that the disk either becomes gravitationally unstable or unrealistically massive compared to the likely mass of the cloud.  From Fig.~2, we see that these conditions are satisfied for cloud column densities  $\la10^{23.5}\ut cm -2 $, for black hole masses approaching $10^8\msol$ the  disk may be truncated because the entire cloud is captured.  For larger cloud column densities the disk may be gravitationally unstable and fragment into stars.   Note that these limits on cloud column density are only a rough guide as the disk column density depends quadratically on the cloud speed and the angular momentum cancellation parameter (see eq 7).

\begin{figure} 
\center 
\includegraphics[scale=1,angle=0]{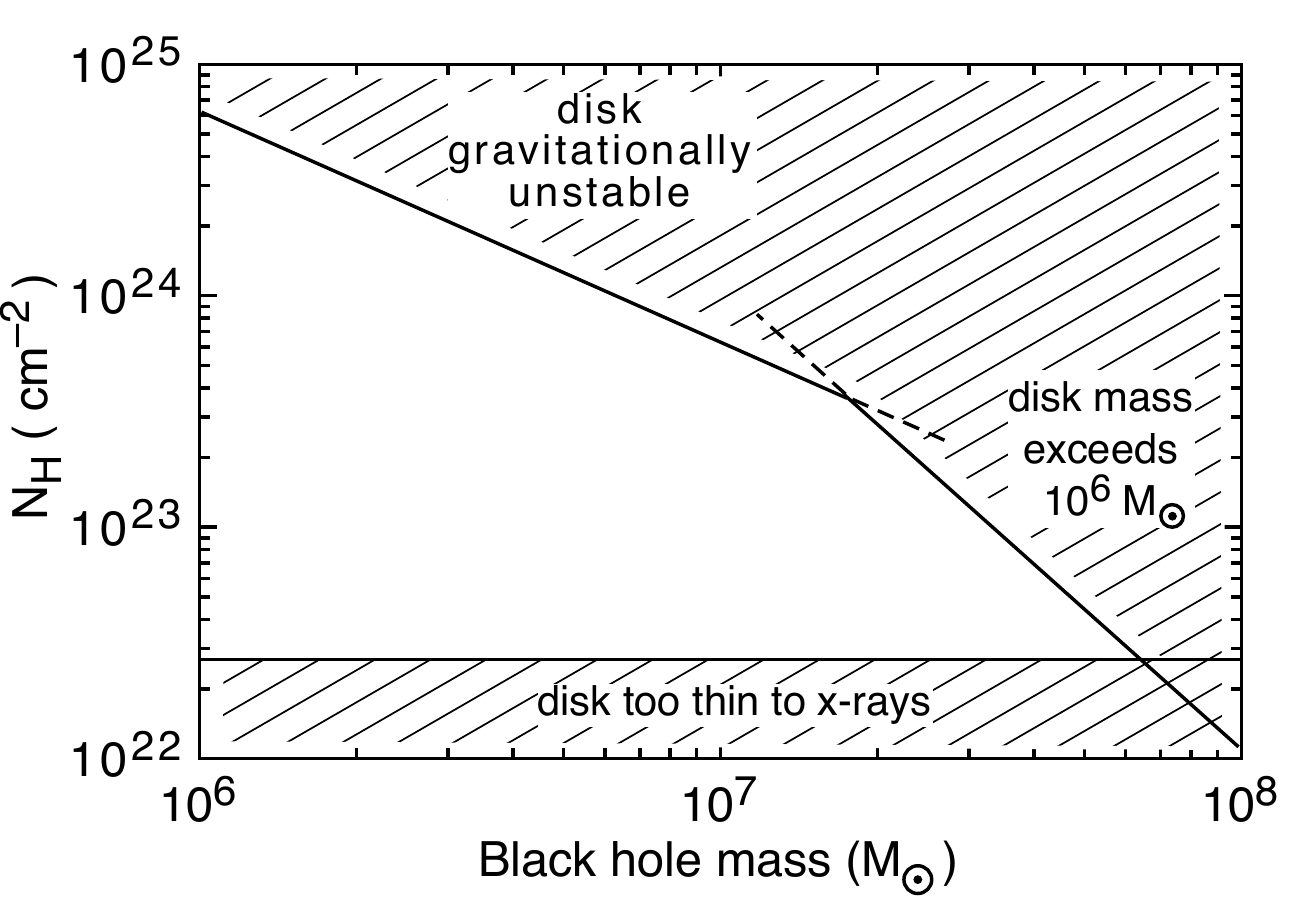} 
\caption{Constraints on the initial hydrogen column density $N_\mathrm{H}$ of the partially captured cloud as a function of the mass of the central black hole. Viable values lie in the unshaded region -- outside of this, the disk formed by capture is either unstable, is too massive, or has insufficient surface density for X-ray irradiation to produce the correct masing conditions (see text). }
\end{figure}

\section{Discussion}

AGN maser disks with accurate black hole mass determinations and Keplerian rotation curves sit well within the sphere of influence of their host black hole and are not self-gravitating.  We showed that the outer radii of the megamaser disks in this sample scale linearly with host black hole mass.  We then considered a scenario for the formation of these disks in which an incoming cloud temporarily engulfs the black hole and is partially captured (Sanders 1981; Bottema \& Sanders 1986; Paper I).  We showed that, for plausible estimates of the mass and angular momentum of the captured material, this process naturally reproduces the empirical linear relationship between maser disk size and black hole mass.   The capture of clouds with column densities $\la 10^{23.5}\ut cm -2 $ results in a non-self-gravitating disks of the correct scale and sufficient column density to allow X-ray irradiation from the central source to reproduce thin megamaser disks.   By contrast, the capture of a cloud with higher column density would instead create a self-gravitating disk giving rise to rapid star formation.  In Paper I we showed that this can explain the recent formation of the compact disks of stars within a fraction of a parsec of the Galactic center black hole Sgr A* (Paper I; Bonnell \& Rice 2008; Mapelli et al.~2008; Alig et al.~2011).   Note that the transition column density is sensitive to the speed of the incoming cloud and the degree of the angular momentum cancellation, though we have chosen plausible values of these parameters.
 
This picture relies on the presence of dense clouds close to the nucleus of galaxies. The circumnuclear molecular ring (e.g.\ Christopher et al. 2005), 1.7\,pc from Sgr A* in our own Galaxy, and the circumnuclear rings found on scales of several parsecs from the center of numerous Seyfert galaxies suggest an ample supply of material.  Recent simulations suggest that gas supply to galactic centers is controlled by angular momentum transfer from one massive gas clump to another during gravitational encounters (Namekata \& Habe 2011).   Some of these inward-moving clouds may interact with their host supermassive black holes.  The rate of migration of molecular material is estimated to  give a potential black hole interaction rate of $\sim 10^{-6}\ut yr -1 $.   Relating these estimates to the occurrence rate of megamaser disks requires estimates of the disk lifetimes and the near edge-on viewing angle needed for maser amplification.  The single-dish detection rate of H$_2$O megamasers in Seyfert 2 galaxies and LINERs is about 15\%, and about 20\% of those have kinematics consistent with orbital motion on sub-parsec scales (Lo 2005).  The estimated maser beaming angle, $\sim 10^\circ$ (e.g.\ Maloney 2002) then implies that $\sim20$\% of all Seyfert 2 galaxies possess similar disks.  The VLBI follow-up results suggest that roughly half may be thin, Keplerian disks.  The ordered kinematics of these disks suggest lifetimes of hundreds of orbital periods or more; the orbital period at 0.3\,pc from a $10^7\,\msun$ black hole is $\sim 3 \ee 3 \u yr $ implies lifetimes in excess of $10^6\u yr $, consistent with the time scale needed to  warp the disk via resonant relaxation (Alexander \& Bregman 2009, 2011).  The formation of gravitationally-unstable disks is likely to be as common simply because molecular clouds in the inner regions of galaxies tend to have column densities $\ga 10^{24}\ut cm -2 $.  These transient disks may also host megamasers (Milosavljevi\'c \& Loeb 2004), so that only a fraction of megamaser AGN may have disks that are very close to Keplerian enabling accurate black hole mass determinations.

Finally, we note that the partial capture of a cloud imparts an impulse to the black hole / disk system of approximately $M_\mathrm{disk} v$, where $v\sim 200\kms $ is the incident cloud velocity.  For plausible parameters this gives recoil velocities of $\sim 10$--$20 \kms$;  while this is potentially detectable, the recoil will be rapidly damped by dynamical friction on the surrounding stars, on a timescale of the stellar crossing time across the sphere of influence, $\sim 10^3$\,years.

\acknowledgments
This research has been supported by the Australian Research Council through Discovery Project grant DP 0986386 and the National Science Foundation through grant AST-0807400.

\end{document}